%% file: ms.tex
\documentclass[preprint2]{aastex}
\shorttitle{Serpens SMM 1 Millimeter Imaging}
\shortauthors{Choi}

\usepackage{times}
\slugcomment{To appear in the Astrophysical Journal}

\begin{document}

\fontsize{10}{10.6}\selectfont

\title{Multiplicity of the Protostar Serpens SMM 1
       Revealed by Millimeter Imaging}
\author{\sc Minho Choi}
\affil{International Center for Astrophysics,
       Korea Astronomy and Space Science Institute,
       Daedukdaero 838, Yuseong, Daejeon 305-348, South Korea;
       minho@kasi.re.kr.}

\begin{abstract}

\fontsize{10}{10.6}\selectfont

The Serpens SMM 1 region was observed in the 6.9 mm continuum
with an angular resolution of about 0\farcs6.
Two sources were found to have steep positive spectra
suggesting emission from dust.
The stronger one, SMM 1a,
is the driving source of the bipolar jet known previously,
and the mass of the dense molecular gas traced by the millimeter continuum
is about 8 $M_\odot$.
The newly found source, SMM 1b, positionally coincides
with the brightest mid-IR source in this region,
which implies that SMM 1b is yet another young stellar object.
SMM 1b seems to be less deeply embedded than SMM 1a.
SMM 1 is probably a protobinary system
with a projected separation of 500 AU.
\end{abstract}


\section{INTRODUCTION}

The Serpens dark cloud is a nearby star-forming region
(see Harvey et al. 2007 and references therein).
The cloud core contains a cluster of low-mass young stellar objects,
and some of them are extremely young protostars
(Testi \& Sargent 1998; Harvey et al. 2007; Winston et al. 2007).
The core also harbors several Herbig-Haro objects and molecular outflows
(Davis et al. 1999; Hodapp 1999).
The most luminous and most deeply embedded object among them is SMM 1
(Harvey et al. 1984; Casali et al. 1993; Enoch et al. 2007).
SMM 1 is a Class 0 source and associated with a bipolar jet/outflow
(Rodr{\'\i}guez et al. 1989; Curiel et al. 1996;
White et al. 1995; Larsson et al. 2000).

Because of the extremely high extinction,
the small scale structure of SMM 1 has been accessible
only with radio interferometric observations.
Images of centimeter continuum revealed a well-collimated bipolar jet
showing a large proper motion away from the central protostar,
indicating a dynamical age of $\sim$60 yr
(Rodr{\'\i}guez et al. 1989; Curiel et al. 1993).
Hogerheijde et al. (1999) imaged SMM 1 in the millimeter continuum
and found a single compact source
surrounded by an envelope of $\sim$9 $M_\odot$.
Hogerheijde et al. (1999) also found
that the molecular outflow driven by SMM 1 has a complicated structure.

One of the interesting issues in the study of star formation
is the multiplicity,
because the majority of stars belong to multiple-star systems
(Duquennoy \& Mayor 1991).
The multiplicity of young pre-main-sequence stars
is higher than that of main-sequence stars (Duch{\^e}ne et al. 2007),
which implies that multiple systems
form in the early stage of star formation.
In fact, bright protostars,
such as IRAS 16293--2422 and NGC 1333 IRAS 4A,
are usually found to be multiple systems
when imaged with a high angular resolution (Looney et al. 2000).
Serpens SMM 1 has been considered
as an interesting example of an isolated protostar
that can be understood well
because it is relatively bright, massive, and simple.
However, there were some indications of complexity,
such as the near-IR excess (Larsson et al. 2000).
Therefore, it is necessary to investigate the structure of SMM 1
with a high angular resolution and a high sensitivity.

In this paper, we present the results of
our observations of the Serpens SMM 1 region
in the 6.9 mm continuum with the Very Large Array (VLA)
of the National Radio Astronomy Observatory.
We describe our radio continuum observations and archival data in Section 2.
In Section 3 we report the results of the continuum imaging.
In Section 4 we discuss the star-forming activities in the SMM 1 region.

\section{OBSERVATIONS}

The Serpens SMM 1 region was observed using VLA
in the standard Q-band continuum mode (43.3 GHz or $\lambda$ = 6.9 mm).
Twenty-three antennas were used
in the D-array configuration on 2001 December 22,
and twenty-six antennas were used
in the C-array configuration on 2004 March 27.
The phase-tracking center was
$\alpha_{2000}$ = 18$^{\rm h}$29$^{\rm m}$49\fs65,
$\delta_{2000}$ = 01\arcdeg15$'$20\farcs6.
The phase was determined
by observing the nearby extragalactic source 1851+005 (G33.50+0.20).

In the D-array observations, the flux was calibrated
by setting the flux density of the quasar 1331+305 (4C 30.26) to 1.18 Jy,
which is the flux density measured within a day of our observations
according to the VLA Calibrator Flux Density Database.
Comparison of the amplitude gave a flux density of 0.65 Jy for 1851+005.
In the C-array observations, the quasar 0410+769 (4C 76.03)
was observed for the purpose of flux calibration,
but its flux around the time of our observations
is not listed in the database.
Since the flux of 0410+769 varies between 0.25 and 0.55 Jy
with a variability time scale of about a month,
it could not be used for the calibration.
The flux of 1851+065 is much more stable than 0410+769,
and the flux scale of the C-array data was calibrated
by setting the flux density of 1851+005 to 0.65 Jy,
as found in the D-array observations.
The flux scale was checked by comparing the total flux densities of SMM 1
from the two observing tracks.
In this comparison, the D-array dataset was limited
to make the shortest {\it uv} length
as the same as that of the C-array dataset.
The resulting flux densities of SMM 1 agreed within 1\%.
Therefore, the flux scale of the C-array data
is consistent with that of the D-array data.

\subsection{Archival Data}

In order to investigate the source structure and the continuum spectra,
we analyzed several VLA datasets
of the X-band continuum (8.5 GHz or $\lambda$ = 3.5 cm)
and C-band continuum (4.9 GHz or $\lambda$ = 6.2 cm) observations
retrieved from the NRAO Data Archive System.
The data were calibrated and imaged following a standard procedure.
The NRAO observing project AC 504 includes data
from the A-array observations made on 1998 May 9, May 30, and June 1.
The 3.5 cm map produced by us is consistent with that of Raga et al. (2000).
However, the flux densities of the outflow sources
are much weaker than what were seen eight years before (Curiel et al. 1993).
Therefore, to measure the flux densities,
we also analyzed the dataset of the NRAO observing project AC 563,
which includes data 
from the A-array observations made on 2000 October 27
(about 1.2 yr before our 6.9 mm observations in D-array).
While the maps from the AC 504 dataset were more useful
in identifying the sources in the 6.9 mm maps,
the flux densities from the AC 563 dataset were used
in the analysis of source spectra.

\section{RESULTS}

\input{fig1.tex}

Figure 1$a$ shows the 6.9 mm continuum map of the SMM 1 region
from the observations in 2001.
The central source (source 1) was clearly detected
and has an extrusion toward the northwest.
Examinations of the CLEAN components showed
that this extrusion is well separated from source 1.
To see it clearly, a map was made with a restoring beam
smaller than the usual synthesized beam (Fig. 1$b$).
While details of this ``super-resolution'' map
should not be overinterpreted,
it shows that the extrusion is an unresolved source (source 2)
located close to the outflow axis and associated with the outflow knot E.
Source 1 seems to be elongated in the northeast-southwest direction.

\input{fig2.tex}

\input{tab1}

Figure 2$a$ shows the 6.9 mm continuum map from the observations in 2004.
For source 1, the peak position of the 6.9 mm source
agrees with that of the 3.5 cm source within 0\farcs1
as well as with those of the H$_2$O and OH masers
present in the region to within 1$''$
(Rodr{\'\i}guez et al. 1989; Moscadelli et al. 2006).
The position of the millimeter source given by Hogerheijde et al. (1999)
is $\sim$2\farcs2 west of the 6.9 mm source position.
The reason for this disagreement is not clear.
We presume that source 1 detected by us
and the source detected by Hogerheijde et al. (1999)
are actually the same object.
Source 2 of the 2001 epoch was undetected in the map of the 2004 epoch.
If its flux were unchanged,
it should have been detected at a 10$\sigma$ level,
where $\sigma$ is the rms noise.
Therefore, source 2 must be time variable.
Source 3 is associated with a 3.5 cm continuum source.
Though Curiel et al. (1993) named it knot F,
it is not exactly located on the outflow axis.
We will discuss the nature of this source in Section 4.2.

Figure 2$b$ shows the map from the D- and C-array datasets combined.
The bright and compact central source
is surrounded by an extended emission structure.
The extended source is clumpy
and elongated in the northeast-southwest direction
with a full length of $\sim$5$''$.
The flux of source 2 in Figure 2$b$ is not very meaningful
because it is time variable.

Table 1 lists the parameters of the 6.9 mm continuum sources.
In the discussion below,
sources 1, 2, and 3 will be referred to
as SMM 1a, knot E, and SMM 1b, respectively.

\section{DISCUSSION}

\subsection{SMM 1a}

\input{fig3.tex}

Comparison of the centimeter and millimeter continuum maps clearly shows
that SMM 1a is the driving source of the bipolar radio jet.
To understand the origin of the 6.9 mm emission,
the spectral energy distribution (SED) of SMM 1a was constructed
from flux measurements with radio interferometric observations (Fig. 3).
The 6.2 cm, 3.5 cm, and 6.9 mm flux densities are from this work,
and the 3.4/3.2 and 2.7 mm values are from Hogerheijde et al. (1999).
The 1.4 mm flux of Hogerheijde et al. (1999) was not included in the plot
because the minimum {\it uv} length of the observations
was a few times larger than those of the other observations,
which may result in resolving out contributions from large-scale structures.
The uncertainties shown in Figure 3 include
the contribution from the uncertainties of the absolute flux scale (2--10\%).

In the centimeter-millimeter wavelength range,
the SED of a low-mass young stellar object usually has two components:
dust emission from a disk/envelope system
with a large ($\gtrsim 2$) spectral index
and free-free emission from a thermal radio jet
with a small ($\lesssim 1$) spectral index
(Reynolds 1986; Anglada et al. 1998). 
The SED of SMM 1a was fitted with a sum of two power-law components,
$F_\nu = F_1 \nu^{\alpha_1} + F_2 \nu^{\alpha_2}$,
where $F_\nu$ is the flux density, $\nu$ is the frequency,
and $\alpha$ is the spectral index.
The best-fit spectrum has spectral indices
of $\alpha_1$ = 3.82 $\pm$ 0.11 and $\alpha_2$ = --0.02 $\pm$ 0.14
and suggests that about 90\% of the 6.9 mm flux comes from dust.

With the SED of dust component, the mass of molecular gas can be estimated.
The mass emissivity given by Beckwith \& Sargent (1991) is assumed,
\begin{equation}
   \kappa_\nu = 0.1 \left({\nu\over{\nu_0}}\right)^\beta
                {\rm cm^2\ g^{-1}},
\end{equation}
where $\nu_0$ = 1200 GHz, and $\beta$ is the opacity index
($\beta \approx \alpha - 2$, in the millimeter range).
If the emission is optically thin,
the mass can be estimated by
\begin{equation}
   M = {{F_\nu D^2}\over{\kappa_\nu B_\nu(T_d)}},
\end{equation}
where $D$ is the distance to the source, $B_\nu$ is the Planck function,
and $T_d$ is the dust temperature.

\input{fig4.tex}

Assuming $D$ = 260 pc and $T_d$ = 36 K
(Strai{\v z}ys et al. 1996; Larsson et al. 2000),
the mass of SMM 1a is $M$ = 8 $\pm$ 3 $M_\odot$,
which includes the inner protostellar envelope and the disk.
If the dust is colder (see Section 4.2), the mass can be larger.
For comparison,
the mass derived by Hogerheijde et al. (1999) is $\sim$4 $M_\odot$
(when scaled to the same distance),
and the mass of the envelope within an 80$''$ beam is 15.4 $M_\odot$
(Enoch et al. 2007).
The large mass of the disk/envelope and the powerful thermal radio jet
corroborate the suggestion
that SMM 1a is an actively accreting Class 0 protostar.

\subsection{SMM 1b}

SMM 1b is located 1\farcs8 (470 AU at a distance of 260 pc)
away from SMM 1a,
and they were clearly separated only in the C-array map of the 2004 epoch.
It has a steeply rising spectrum ($\alpha \approx 2.04$)
between 3.5 cm and 6.9 mm (Fig. 3).
The SED suggests
that a significant part of the 6.9 mm flux may be from dust,
which would imply that SMM 1b is not an outflow knot
but a dense core of molecular gas.
If the mass is simply scaled from SMM 1a
using the flux ratio as a very rough estimate,
the mass of the molecular gas associated with SMM 1b
is $\sim$0.7 $M_\odot$.

The {\it Spitzer} images provide further useful information (Fig. 4).
While the 4.5--8.0 $\mu$m images
show the outflow in the overall structure (Harvey et al. 2007),
the brightest mid-IR source (source 141) coincides with SMM 1b,
and SMM 1a was undetected.
The peak position of the 24 $\mu$m source
is situated in the middle between SMM 1a and 1b.
In other words,
SMM 1b dominates the mid-IR part of the SED,
SMM 1a dominates in the millimeter range,
and they may be comparable in the far-IR flux.

As shown in Figure 1$b$,
SMM 1b is comparable to other clumpy structures around SMM 1a
(e.g., the clump 1\farcs5 northeast of SMM 1a) in terms of intensity.
However, SMM 1b is different from the other clumps in several respects.
First, only SMM 1b was clearly detected in the C-array map (Fig. 1$a$),
which suggests that SMM 1b has a compact structure.
Second, SMM 1b has counterparts in other wavelengths:
knot F in the centimeter continuum images and source 141 in the IR images.
Third, the other clumps seem to belong to an extended structure
elongated in the direction perpendicular to the bipolar jet,
which suggests that it is a disk-like flat structure around SMM 1a,
while SMM 1b is located away from this extended structure.
The most likely interpretation is
that SMM 1b is yet another young stellar object in this region.
In other words, SMM 1 may be a protobinary system.

The SMM 1a/b system illustrates the importance of high-resolution imaging.
The SED of SMM 1 in previous studies (e.g., Larsson et al. 2000)
is actually a combination of the SEDs of the two sources.
While an accurate SED of each source is not available
unless the system can be resolved in the far-IR band,
the dust temperature of SMM 1a may be lower
than what was derived previously.
SMM 1b is probably less deeply embedded than SMM 1a
but still very young
as it was not clearly detected in the 3.6 $\mu$m band
and undetected in shorter wavelengths.

In the classification scheme of multiple young stellar objects
using the binary separation, as suggested by Looney et al. (2000),
SMM 1 belongs to the common-envelope type.
It will be interesting to image the system with a higher angular resolution
to see the relation between the two binary components,
such as the alignment of the disks.

It is interesting to note
that there is an H$_2$ jet that seems to be driven by SMM 1 (Hodapp 1999).
It can also be seen in the IR images (Fig. 4).
Since the direction of the H$_2$ jet has a position angle different
from that of the bipolar radio jet by $\sim$25\arcdeg,
the two jets may have separate driving sources.
SMM 1b is a probable driving source of the H$_2$ jet.

\subsection{Knot E}

Knot E has a flat ($\alpha \approx 0.09$) spectrum (Fig. 3),
which suggests that the 6.9 mm flux
is mostly free-free emission from the thermal jet.
The nondetection at the 2004 epoch suggests that knot E is fading rapidly.
Such a fading has been observed in other radio jets
such as the HH 80--81/GGD 27 jet (Mart{\'\i} et al. 1998).

\subsection{VLA 8}

VLA 8 is a centimeter continuum source
located 23$''$ northeast of SMM 1 (Eiroa et al. 2005).
It was detected in the 6.2 and 3.5 cm images as an unresolved source
with flux densities of 0.481 $\pm$ 0.010 and 0.442 $\pm$ 0.011 mJy,
respectively.
The peak position is
$\alpha_{2000}$ = 18$^{\rm h}$29$^{\rm m}$51\fs04,
$\delta_{2000}$ = 01\arcdeg15$'$33\farcs9.
VLA 8 was undetected at 6.9 mm ($<$ 0.24 mJy beam$^{-1}$).
The SED suggests
that the centimeter continuum flux comes from free-free emission
and that there is no substantial emission from dust.
The nature of VLA 8 is not clear.

\acknowledgements

We thank Jeong-Eun Lee and Miju Kang
for helpful discussions and suggestions.
This work was supported by the National Research Foundation of Korea
through the grant R01-2007-000-20196-0.
The National Radio Astronomy Observatory is
a facility of the National Science Foundation
operated under cooperative agreement by Associated Universities, Inc.
This work is based in part on observations
made with the {\it Spitzer} Space Telescope,
which is operated by the Jet Propulsion Laboratory,
California Institute of Technology, under a contract with NASA.

%\TimeStamp

\end{document}

%% file: fig1.tex
\begin{figure}[!t]
\epsscale{1}
\plotone{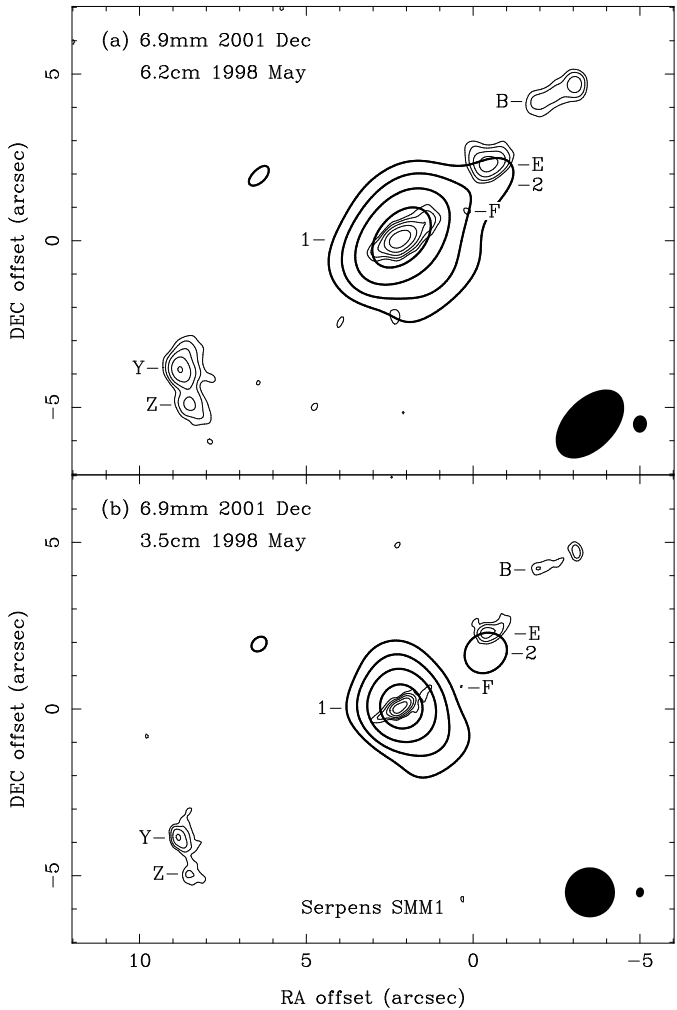}
\caption{\small\baselineskip=0.825\baselineskip
Uniform-weight maps of the 6.9 mm continuum ({\it thick contours})
toward the Serpens SMM 1 region from the observations in 2001 December.
The rms noise is 0.2 mJy beam$^{-1}$.
The contour levels are 1, 2, 4, and 8 times 0.6 mJy beam$^{-1}$.
The 6.9 mm sources are labeled.
({\it a})
Map restored with the usual synthesized beam
shown in the bottom right corner:
FWHM = 2\farcs5 $\times$ 1\farcs5 and P.A. = --45\arcdeg.
{\it Thin contours}:
Map of the 6.2 cm continuum from the AC 504 dataset.
The rms noise is 0.012 mJy beam$^{-1}$.
The contour levels are 1, 2, 4, 8, and 16 times 0.04 mJy beam$^{-1}$.
The synthesized beam is
FWHM = 0\farcs51 $\times$ 0\farcs42 and P.A. = --5\arcdeg.
({\it b})
Map restored with a 1\farcs5 circular beam shown in the bottom right corner.
{\it Thin contours}:
Map of the 3.5 cm continuum from the AC 504 dataset (Raga et al. 2000).
The rms noise is 0.009 mJy beam$^{-1}$.
The contour levels are 1, 2, 4, 8, and 16 times 0.04 mJy beam$^{-1}$.
The synthesized beam is
FWHM = 0\farcs31 $\times$ 0\farcs24 and P.A. = --10\arcdeg.
Outflow knots are labeled
following the convention used by Curiel et al. (1993).}
\end{figure}

%% file: fig2.tex
\begin{figure}[!t]
\epsscale{1}
\plotone{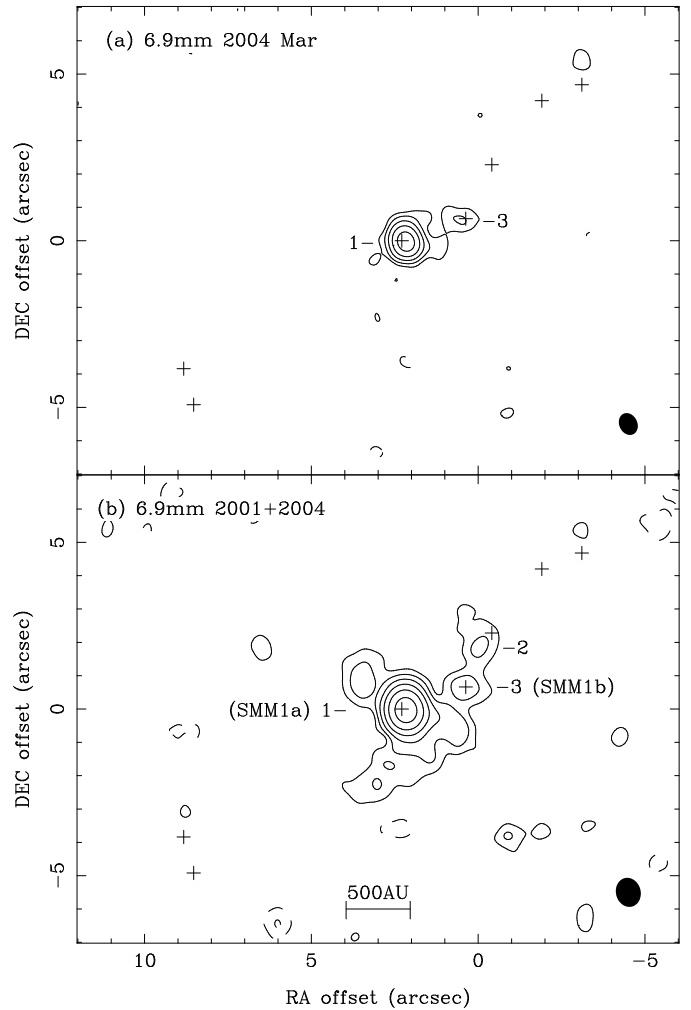}
\caption{\small\baselineskip=0.825\baselineskip
Natural-weight maps of the 6.9 mm continuum toward the SMM 1 region.
The 6.9 mm sources are labeled.
({\it a})
Map from the observations in 2004 March.
The rms noise is 0.09 mJy beam$^{-1}$.
The contour levels are 1, 2, 4, 8, and 16 times 0.3 mJy beam$^{-1}$.
Dashed contours are for negative levels.
Shown in the bottom right corner is the synthesized beam:
FWHM = 0\farcs69 $\times$ 0\farcs54 and P.A. = 26\arcdeg.
({\it b})
Map from the 2001 and 2004 datasets combined.
The rms noise is 0.05 mJy beam$^{-1}$.
The contour levels are 1, 2, 4, 8, 16, and 32 times 0.15 mJy beam$^{-1}$.
Shown in the bottom right corner is the synthesized beam:
FWHM = 0\farcs88 $\times$ 0\farcs75 and P.A. = 15\arcdeg.
The straight line near the bottom
corresponds to 500 AU at a distance of 260 pc.
{\it Plus signs}:
Position of the 3.5 cm sources shown in Fig. 1$b$.}
\end{figure}

%% file: tab1.tex
\begin{deluxetable}{llcccc}
\tabletypesize{\small}
\tablecaption{Millimeter Continuum Sources in the Serpens SMM 1 Region}%
\tablewidth{0pt}
\tablehead{
\colhead{Source} & \colhead{Name} & \multicolumn{2}{c}{Peak Position}
& \colhead{Peak Flux} & \colhead{Total Flux} \\
\cline{3-4}
& & \colhead{$\alpha_{\rm J2000.0}$} & \colhead{$\delta_{\rm J2000.0}$}
& \colhead{(mJy beam$^{-1}$)} & \colhead{(mJy)}}%
\startdata
1 & SMM 1a & 18 29 49.80 & 01 15 20.6
  & 7.67     $\pm$ 0.05    &    13.1\phn\ $\pm$ 0.4\phn \\
2 & knot E & 18 29 49.63 & 01 15 22.3
  & 0.9\phn\ $\pm$ 0.2\phn & unresolved \\
3 & SMM 1b & 18 29 49.69 & 01 15 21.2
  & 0.47     $\pm$ 0.05    & \phn1.10     $\pm$ 0.09 \\
\enddata\\
\tablecomments{
Units of right ascension are hours, minutes, and seconds,
and units of declination are degrees, arcminutes, and arcseconds.
Flux densities at 6.9 mm are corrected for the primary beam response.
For SMM 1a and 1b, the flux densities
are from the D- and C-array data combined (Fig. 2$b$).
For knot E, it is from the D-array data only (Fig. 1$b$).
Total flux of each source was measured
in a circumscribed rectangle of 2$\sigma$ contour.}%
\end{deluxetable}

%% file: fig3.tex
\begin{figure}[!t]
\epsscale{1}
\plotone{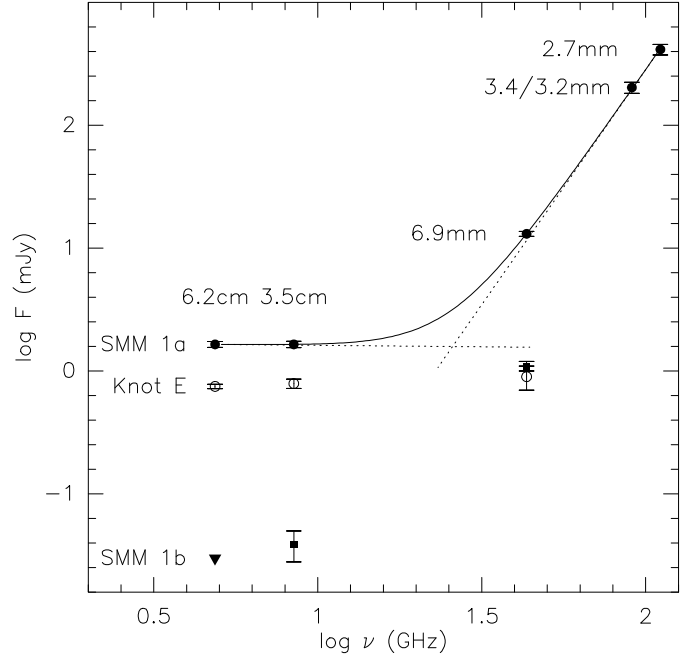}
\caption{\small\baselineskip=0.825\baselineskip
Spectral energy distribution of SMM 1a ({\it filled circles}),
SMM 1b ({\it filled squares}), and knot E ({\it open circles}).
{\it Solid curve}:
Best-fit two-component power-law spectrum of SMM 1a.
{\it Dotted straight lines}:
Each component of the fit.
{\it Filled triangle}:
The upper limit (3$\sigma$) for SMM 1b at 6.2 cm.}
\end{figure}

%% file: fig4.tex
\begin{figure*}[!t]
\epsscale{1.8}
\plotone{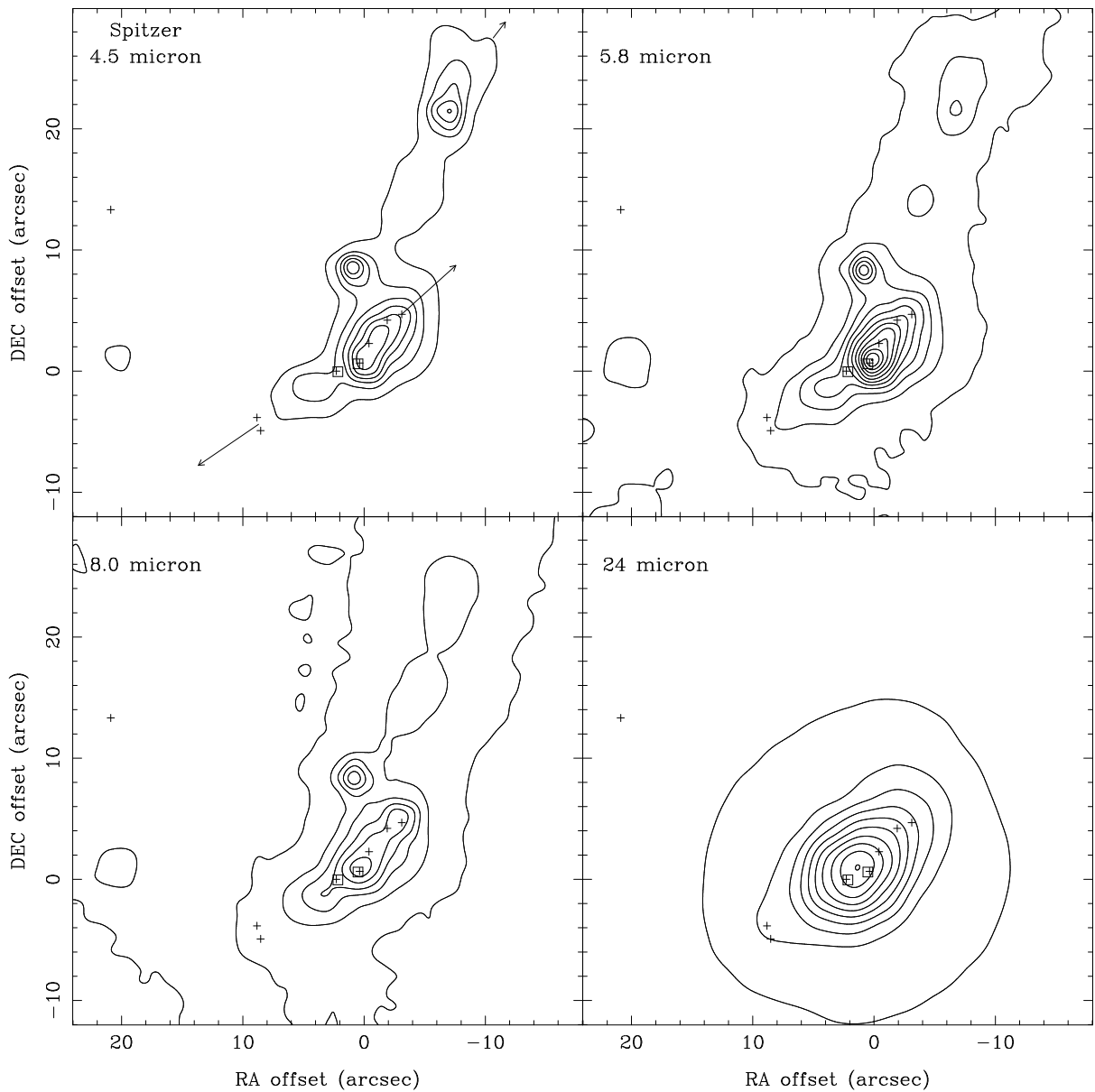}
\caption{\small\baselineskip=0.825\baselineskip
Images of the 4.5, 5.8, 8.0, and 24 $\mu$m bands
toward the SMM 1 region from the {\it Spitzer} data archive.
The contour levels are in arbitrary linear sequences.
See Harvey et al. (2007) for the flux scale.
{\it Squares}:
Position of the 6.9 mm sources SMM 1a/b.
{\it Plus signs}:
Position of the 3.5 cm sources shown in Fig. 1$b$
and VLA 8 (near the eastern edge).
{\it Long arrows}:
Proper motion vectors
of the bipolar radio jet (0\farcs12 yr$^{-1}$; Curiel et al. 1993).
{\it Short arrow}:
Proper motion vector of the H$_2$ jet
($\sim$0\farcs03 yr$^{-1}$; Hodapp 1999).}
\end{figure*}